\documentclass[a4paper,12pt]{article}
\usepackage{graphicx}
\usepackage{graphics}
\usepackage{amsmath,amsfonts,amssymb}
\usepackage{overcite}

\renewcommand{\baselinestretch}{2}
\begin{document}

\section{Copyright}

Copyright 2009 American Institute of Physics. This article may be downloaded for personal use only. Any other use requires prior
permission of the author and the American Institute of Physics. The following article appeard in J.\ Chem.\ Phys.\, \textbf{131}, 044312 (2009)  and may be found at\\
 http://jcp.aip.org/resource/1/jcpsa6/v131/i4/p044312\_s1

\title{Local softness, softness dipole and polarizabilities of functional
groups: application to the side chains of the twenty amino acids}
\author{Alisa Krishtal$^1$, Patrick Senet$^2$\thanks{%
psenet@u-bourgogne.fr}, Christian Van Alsenoy$^1$}
\date{}
\maketitle

\begin{abstract}
\noindent The values of molecular polarizabilities and softnesses of the
twenty amino acids were computed ab initio (MP2). By using the iterative
Hirshfeld scheme to partition the molecular electronic properties, we
demonstrate that the values of the softness of the side chain of the twenty
amino acid are clustered in groups reflecting their biochemical
classification, namely: aliphatic, basic, acidic, sulfur containing, and
aromatic amino acids . The present findings are in agreement with previous
results using different approximations and partitioning schemes [P. Senet
and F. Aparicio, J. Chem. Phys. 126,145105 (2007)]. In addition, we show
that the polarizability of the side chain of an amino acid depends mainly on
its number of electrons (reflecting its size) and consequently cannot be
used to cluster the amino acids in different biochemical groups, in contrast
to the local softness. Our results also demonstrate that the global softness
is not simply proportional to the global polarizability in disagreement with
the intuition that ``a softer moiety is also more polarizable''. Amino acids
with the same softness may have a polarizability differing by a factor as
large as 1.7. This discrepancy can be understood from first principles as we
show that the molecular polarizability depends on a ``softness dipole
vector'' and not simply on the global softness.
\end{abstract}

\noindent $^1$ Chemistry Department, University of Antwerp,
Universiteitsplein 1, B2610 Antwerp, Belgium\newline
\noindent $^2$ Institut Carnot de Bourgogne, UMR 5209 CNRS, Universit\'e de
Bourgogne, 9 Avenue Alain Savary BP 47870, F-21078 Dijon Cedex, France

\vspace{2cm}

\newpage

\section{Introduction}

Classification of molecules in families depending on their functional groups
is as old as chemistry itself. In principle, quantum mechanical calculations
contain the necessary information to evaluate the chemical (quantum)
similarity of any two molecules or any two fragments within a set of
different molecules. Therefore \emph{ab initio} calculations should be able
to recover the empirical classifications of chemistry and more generally to
predict the ``similarity'' between two molecules in a chemical library. The
later application is important in pharmacology where \emph{ab initio}
calculations can be used to discriminate a potential drug within a large set
of molecules.\cite{QuantumSimilarity.ref}

In the present work, one shows how certain descriptors defined in Density
Functional Theory (DFT) can be used to discriminate a cluster of molecules
within a ``family'', where a family is defined as a set of molecules having
a fragment in common. Two members of such a family differ from each other by
a ``variable fragment''. For example, the amino acids form a family because
each member has a fragment in common with the others (the amino-acid part or
backbone) and a variable fragment (side chain). One confirms below that the
concepts of local softness\cite{softness.ref} $s(\mathbf{r)~}$ allow to
identify different subgroups of ``similar'' molecules within a family.

The local softness $s(\mathbf{r})$ can be computed at any point in space and
is proportional to the change in the electronic density $\rho (\mathbf{r})$
induced by a shift $\delta N$ in the number of electrons $N$ of a molecule.%
\cite{Parrbook.ref,softness.ref} The integration of the local softness of an
isolated molecule over all the space is the (global) molecular softness $%
S=\int d\mathbf{r~}s(\mathbf{r}).$ The molecular hardness is defined as the
inverse of (global) softness $\eta =1/S$.\cite{softness.ref,Parrbook.ref}
Both softness and hardness have been extensively studied in recent years.%
\cite%
{Geerlings1.ref,Geerlings2.ref,Kulkarni.ref,Newbook.ref,Pearson86.ref,Hemelsoet.ref,Ayers.ref,Cohen.ref,Cardenas.ref,Kar.ref,ChemREV.ref,ChermetteREV.ref}

Local softness can be computed by using the following relation\cite%
{softness.ref,Parrbook.ref}

\begin{equation}
s(\mathbf{r})=S~\left[ \frac{\delta \mu }{\delta v(\mathbf{r})}\right] _{N},
\label{local-s.eq}
\end{equation}%
where $\mu $ is the molecular chemical potential and $v(\mathbf{r})$ the
external potential. The functional derivative in Eq.\ (\ref{local-s.eq}) is
evaluated at constant\ $N$. In the present paper, one applies the "frozen
orbital approximation" to evaluate the derivative in Eq.\ (\ref{local-s.eq})
and one chooses the chemical potential $\mu ~$as the average between the
energies of the highest occupied (HOMO) and lowest unoccupied (LUMO)
molecular orbitals,\cite%
{ChemREV.ref,ChermetteREV.ref,Parrbook.ref,DFT97.ref,CPL.ref} i.e. $\mu
=\left( \epsilon _{LUMO}+\epsilon _{HOMO}\right) /2$.$~$\ The hardness $\eta
~$is defined as the difference between the energies of the frontier
orbitals, also named the HOMO-LUMO gap, i.e. $\eta =1/S=\epsilon
_{LUMO}-\epsilon _{HOMO}$.\cite{Pearson86.ref} Within this approximation,
one has the usual relation 
\begin{equation}
s(\mathbf{r})=\frac{\left[ \rho _{LUMO}(\mathbf{r})+\rho _{HOMO}(\mathbf{r})%
\right] }{2\left[ \epsilon _{LUMO}-\epsilon _{HOMO}\right] }.
\end{equation}%
For an isolated molecule, the spatial variation of $s(\mathbf{r})$ is
entirely due to the variations of the HOMO/LUMO\ frontier orbitals because
the HOMO-LUMO gap remains constant. On the contrary, if one wishes to
compare the values of $s(\mathbf{r})$ between two members of the same family
also the difference in their HOMO-LUMO gaps is of importance.

The variations of $s(\mathbf{r})$ within a molecule, or its comparison
between two molecules, are more easily analyzed by using a coarse-grained
representation of this function obtained by partitioning the electronic
density into fragments.\cite{condens.ref,CoulombHole1.ref,CoulombHole.ref}
Within a fragment $k$, the function $s(\mathbf{r})$ is replaced by a single
value $s_{k}$, called the \textquotedblleft condensed
softness\textquotedblright \cite{condens.ref} and computed by using one of
the partitioning schemes of the electronic properties.\cite%
{Bader.ref,Hirshfeld.ref} In a recent work, one of us demonstrates that such
condensed softness is correctly described as a polarization of the fragment $%
k$ by an effective potential produced by the rest of the molecule.\cite%
{CoulombHole.ref,CoulombHole1.ref} The concept of Coulomb hole, recently
introduced, is a measure for the amount of charge $q_{g}^{h}$ induced by one
fragment on the another. More precisely, for a family with a common fragment 
$1$ and a variable fragment $2$, one demonstrated that the softness$~S$ of
any molecule of the family is related to the softness of its functional
group ($s_{2}$) by a linear relation:\cite{CoulombHole.ref} 
\begin{equation}
S=s_{2}[1-q_{g}^{h}(1)]+\overline{S_{1}},  \label{eq:linear}
\end{equation}%
where $q_{g}^{h}(1)$ defines the average Coulomb hole of the molecular
family and has the meaning of an \textquotedblleft induced
charge\textquotedblright\ within fragment $2$ by one hole $($charge +$e)$ on
fragment $1$ [See Eq.\ (13) in Ref.\ (\ref{CoulombHole1.ref})]. Eq.\ (\ref%
{eq:linear}) allows to compute the average softness $\overline{S_{1}}$ of
the common fragment of a family of molecules and allows to evaluate the
similarities and differences between the individual members of the family:
for the amino acids,\ the positions of the molecules with similar functional
groups were found close to each other on the $(S,s_{2})$ map.\cite%
{CoulombHole.ref}

Another important molecular descriptor of the electronic properties of a
molecule is its polarizability. Softness and polarizability are assumed to
be related: \textquotedblleft a soft species is also more
polarizable\textquotedblright .\cite{Parrbook.ref} But how to compare the 
\emph{local} polarizability and the \emph{local} softness (or frontier
orbitals) of fragments within a molecule from first principles? To answer
this question, the local polarizability of a molecule is defined (Section
2.1.) as follows: 
\begin{equation}
\alpha _{ij}(\mathbf{r})\equiv -er_{i}\left[ \frac{\partial \rho (\mathbf{r})%
}{\partial E_{j}}\right] _{N},  \label{local-pol.eq}
\end{equation}%
where $i$ and $j$ stand for Cartesian directions and $\mathbf{E}$ is a
uniform electric field applied to the molecule. In Eq.\ (\ref{local-pol.eq})
as well as in all equations below, the vector $\mathbf{r}$ represents the
position relative to the center of mass of the molecule, i.e. the origin of
the Cartesian axis is chosen as the center of mass of the molecule by
convention.

The applied field $\mathbf{E~}$ is derived from an external potential $%
\Delta v(\mathbf{r})=e\sum_{j}r_{j}E_{j}~$($e=1.602~10^{-19}\ C$). \ As
shown in the next Section, the local polarizability tensor is related to the
local softness by 
\begin{equation}
\alpha _{ij}(\mathbf{r})=\left[ \alpha _{ij}^{\mu }(\mathbf{r})-\frac{s(%
\mathbf{r})er_{i}}{S}\int d\mathbf{r}^{\prime }s(\mathbf{r}^{\prime
})er_{j}^{\prime }\right] ,  \label{local soft-pol.eq}
\end{equation}%
where $\alpha _{ij}^{\mu }(\mathbf{r})$ is the local polarizability tensor
at a constant electronic chemical potential. The molecular polarizability is
found by integrating the local polarizablity: 
\begin{equation}
\alpha _{ij}=\int d\mathbf{r~}\alpha _{ij}(\mathbf{r}).  \label{volume.eq}
\end{equation}%
The isotropic part of the molecular polarizability tensor is in obvious
notations (Section 2.1.) 
\begin{eqnarray}
\alpha _{iso} &=&\frac{\alpha _{xx}+\alpha _{yy}+\alpha _{zz}}{3},  \notag \\
&=&\alpha _{iso}^{\mu }-\frac{\left\Vert \int d\mathbf{r}s(\mathbf{r})e%
\mathbf{r}\right\Vert ^{2}}{3S}.  \label{tracepol.eq}
\end{eqnarray}%
The numerator of the last term in Eq.\ (\ref{tracepol.eq}) is the square
length of the \textquotedblleft softness dipole vector\textquotedblright , $~%
\mathbf{D}\equiv \int d\mathbf{r}s(\mathbf{r})e\mathbf{r}$.

The local polarizability tensor $\alpha _{ij}(\mathbf{r})$ can also be
partitioned into fragments. In fact, the partitioning of the polarizability
of a molecule into a sum of polarizabilities of its parts has a long
history. In some sense, the concept of ``local polarizability'' is quite old.%
\cite{teixeira-dias.ref} \emph{Ab initio} calculations have established that
the polarizability (more precisely the trace of the polarizability tensor)
of an atom or of a molecule is proportional to its volume.\cite%
{Gough.ref,Bader.ref} This explains some sucess of the ``additive rule'' of
polarizabilities which consists of summing up the empirical values of the
polarizabilities of the atoms or functional groups of a molecule to estimate
its global polarizability.\cite{lefevre.ref,miller.ref}

In the present study, local softness $s(\mathbf{r})$ and local
polarizability $\alpha _{ij}(\mathbf{r})$ computed \emph{ab initio} were
obtained by applying the Hirshfeld scheme, which is based on a partitioning
of the density of the electrons.\cite%
{Hirshfeld.ref,Rousseau.ref,Bultinck.ref,PolarWater.ref,PolarMethanol.ref}
The partitioning of $s(\mathbf{r})$ computed by applying the Hirshfeld
method is compared with previous results\cite{CoulombHole.ref} where the
Contreras et al.\cite{Contreras.ref} partitioning scheme, based on molecular
orbitals, was applied. It should be emphasized that the present study
provides the most extensive study of polarizabilities and softness of
isolated amino acids and of their side chains. The present work differs from
earlier studies of reactivity descriptors of the subset of the twenty amino
acids, which were mainly devoted to determine the protonation site in the
amino acid region.\cite{AA1.ref,AA2.ref,AA3.ref,AA4.ref}

The paper is organized as follows. The theory and numerical methods are
summarized in the next section. Results are presented and discussed in
Section 3. The paper ends with final concluding remarks in the last Section.

\section{Theory and methods}

\subsection{Local polarizability}

The dipole of a molecule with $M$ atoms is given by 
\begin{equation}
\mathbf{P}\mathbf{\equiv }-e\int d\mathbf{r~r~}\rho (\mathbf{r}%
)+\sum_{A=1}^{M}Z_{A}e\mathbf{R}^{A},  \label{dipole.eq}
\end{equation}%
where $e$ is the elementary charge ($e=1.602~10^{-19}\ C),~Z_{A}~$is the
atomic number of atom $A$ and $\mathbf{R}^{A}$ is its position relative to
the center of mass of the molecule chosen as the origin of the Cartesian
coordinates. The components of the dipole polarizability tensor of the
molecule are given by 
\begin{equation}
\alpha _{ij}=\frac{\partial P_{i}}{\partial E_{j}}\mathbf{=}-e\int d\mathbf{%
r~}r_{i}\frac{\partial \rho (\mathbf{r})}{\partial E_{j}},
\label{polarisability.eq}
\end{equation}%
where $i$ and $j$ stand for Cartesian directions and $\mathbf{E}$ is a
uniform electric field applied to the molecule. Therefore, the local
polarizability of a molecule is defined as follows: 
\begin{equation}
\alpha _{ij}(\mathbf{r})\equiv -er_{i}\frac{\partial \rho (\mathbf{r})}{%
\partial E_{j}}.  \label{LocalPOL.eq}
\end{equation}%
The applied field $\mathbf{E~}$ derives from an external potential $v(%
\mathbf{r})=e\sum_{j}r_{j}E_{j}$. The tensor~$\alpha _{ij}(\mathbf{r})$ is
symmetrical because one can write: 
\begin{eqnarray}
\frac{\partial \rho (\mathbf{r})}{\partial E_{j}} &=&\int d\mathbf{r}%
^{\prime }\left[ \frac{\delta \rho (\mathbf{r})}{\delta v(\mathbf{r}^{\prime
})}\right] _{N}\frac{\delta v(\mathbf{r}^{\prime })}{\delta E_{j}},  \notag
\\
&=&\int d\mathbf{r}^{\prime }\chi _{1}(\mathbf{r},\mathbf{r}^{\prime
})er_{j}^{\prime },  \label{deriv.eq}
\end{eqnarray}%
where $\chi _{1}$ is the so-called linear polarizability kernel \cite%
{Resp.ref} with the property $\int d\mathbf{r}^{\prime }\chi _{1}(\mathbf{r},%
\mathbf{r}^{\prime })=0.$\cite{DFT96.ref} The global polarizability is
obtained by integration of the local polarizability over all the space 
\begin{equation}
\alpha _{ij}=\int d\mathbf{r~}\alpha _{ij}(\mathbf{r})=-\int d\mathbf{r}\int
d\mathbf{r}^{\prime }er_{i}\chi _{1}(\mathbf{r},\mathbf{r}^{\prime
})er_{j}^{\prime }.  \label{Polarisability.eq}
\end{equation}%
The last equality in Eq.\ (\ref{Polarisability.eq}) is a well-known relation
between molecular polarizability and the symmetrical kernel $\chi _{1}$ [See
for instance Ref.\ (\ref{Mahan.ref})].

The relation between the local polarizabity and the local and global
softnesses can be established by using the so-called Berkowitz-Parr relation%
\cite{BP.ref} 
\begin{equation}
\chi _{1}(\mathbf{r},\mathbf{r}^{\prime })=\chi _{1}^{\mu }(\mathbf{r},%
\mathbf{r}^{\prime })+\frac{s(\mathbf{r})s(\mathbf{r}^{\prime })}{S},
\label{Berko-Parr.eq}
\end{equation}%
where $\chi _{1}^{\mu }(\mathbf{r},\mathbf{r}^{\prime })~$is the
polarizability kernel at a constant electronic chemical potential $\mu .$%
\cite{DFT96.ref} One has the relation $\chi _{1}^{\mu }(\mathbf{r},\mathbf{r}%
^{\prime })=-~s(\mathbf{r},\mathbf{r}^{\prime })$ where $s(\mathbf{r},%
\mathbf{r}^{\prime })$ is the so-called softness kernel related to the local
softness by integration: $\ s(\mathbf{r})=\int d\mathbf{r}^{\prime }s(%
\mathbf{r},\mathbf{r}^{\prime }).$\cite{BP.ref,DFT96.ref} Replacing $\chi
_{1}$ in Eq.\ (\ref{deriv.eq}) by the right-hand side of Eq.\ (\ref%
{Berko-Parr.eq}), one obtains in obvious notations for the local
polarizabity [Eq.\ (\ref{LocalPOL.eq})] 
\begin{equation}
\alpha _{ij}(\mathbf{r})=\alpha _{ij}^{\mu }(\mathbf{r})-er_{i}s(\mathbf{r}%
)\int d\mathbf{r}^{\prime }\frac{s(\mathbf{r}^{\prime })}{S}er_{j}^{\prime },
\label{pol-local-soft.eq}
\end{equation}%
and for the global polarizability [Eq.\ (\ref{volume.eq})] 
\begin{eqnarray}
\alpha _{ij} &=&\alpha _{ij}^{\mu }-\frac{\int d\mathbf{r}s(\mathbf{r}%
)er_{i}\int d\mathbf{r}^{\prime }s(\mathbf{r}^{\prime })er_{j}^{\prime }}{S},
\notag \\
&=&\alpha _{ij}^{\mu }-\frac{D_{i}D_{j}}{S},  \label{pol-global-soft.eq}
\end{eqnarray}%
where $D_{i}$ is the component of the softness dipole vector 
\begin{equation}
\mathbf{D}=\int d\mathbf{r}s(\mathbf{r})e\mathbf{r},  \label{def-D.eq}
\end{equation}%
in the Cartesian direction $i$. The isotropic part of the polarizability
tensor is 
\begin{eqnarray}
\alpha _{iso} &=&\frac{\alpha _{xx}+\alpha _{yy}+\alpha _{zz}}{3},  \notag \\
&=&\alpha _{iso}^{\mu }-\frac{\left\Vert \int d\mathbf{r\ }s(\mathbf{r})e%
\mathbf{r}\right\Vert ^{2}}{3S},  \notag \\
&=&\alpha _{iso}^{\mu }-\frac{D^{2}}{3S}.  \label{pol-soft-trace.eq}
\end{eqnarray}%
where $D^{2}$ is the square length of the softness dipole vector. It should
be emphasized that the values of $\alpha _{ij}^{\mu }(\mathbf{r})$ and of
the softness dipole vector, as defined above, are defined for a molecule for
which the center of mass is the origin of the Cartesian coordinates.

Relation (\ref{pol-soft-trace.eq}) is interesting since it shows explicitely
how polarizability is related to softness. A large $S$ means a less negative
contribution to the right-hand side of Eq.\ (\ref{pol-soft-trace.eq}) and a
larger value for the polarizability on the left-hand side of Eq.\ (\ref%
{pol-soft-trace.eq}) if $\alpha _{iso}^{\mu }$ is constant. However, the
polarizability depends also explicitely on the softness dipole $D$. It is
the ratio $D^{2}/3S$ which is the most important quantity relating
polarizability and softness. Since polarizability is positive one must have
the following relation 
\begin{equation}
\alpha _{iso}^{\mu }>\frac{D^{2}}{3S}.  \label{bondary.eq}
\end{equation}%
The quantity $D^{2}/3S$ is a lower bound for the polarizability of any
molecule.

\subsection{Hirshfeld partitioning applied to softness and polarizabilities}

In a previous study\cite{CoulombHole.ref}, each amino acid was separated
into a backbone (fragment 1) and a side chain (fragment 2) by applying the
partitioning method proposed by Contreras et al.\cite{Contreras.ref} In the
present work, we test the dependence of the ($S$,$s_{2}$) map on the
partitioning method by using another partitioning of the electronic
properties: the Hirshfeld-I scheme.\cite{Bultinck.ref} The Hirshfeld-I
scheme allows to write the electronic density of a molecule $\rho (\mathbf{r}%
)$ as a sum of atomic contributions~$\rho _{A}(\mathbf{r})$: 
\begin{equation}
\rho (\mathbf{r})=\sum_{A}\rho _{A}(\mathbf{r})=\sum_{A}\omega _{A}(\mathbf{r%
})\rho (\mathbf{r}),  \label{rho.eq}
\end{equation}%
where the summation is over all atoms $A$ of the molecule. The atomic weight
function $\omega _{A}(\mathbf{r})~$in Eq.\ (\ref{rho.eq}) is built
iteratively from the atomic densities obtained in the previous iteration
until self consistence is reached 
\begin{equation}
\omega _{A}^{n}(\mathbf{r})=\frac{\rho _{A}^{n-1}(\mathbf{r})}{\sum_{B}\rho
_{B}^{n-1}(\mathbf{r})}.  \label{wn.eq}
\end{equation}%
The electronic density of an atom at iteration $n=0$ (initial guess) is the
density of the \emph{isolated} atom. Therefore, the initial value of the
weight function $\omega _{A}^{1}~$is identical to the weight function of the
usual non-iterative Hirshfeld scheme: 
\begin{equation}
\omega _{A}^{1}(\mathbf{r})=\frac{\rho _{A}^{0}(\mathbf{r})}{\sum_{B}\rho
_{B}^{0}(\mathbf{r})}.  \label{w1.eq}
\end{equation}%
The number of electrons of each isolated atom is $N_{A}^{0}=Z_{A}=\int d%
\mathbf{r~}\rho _{A}^{0}(\mathbf{r})$. The weight function in Eq.\ (\ref%
{w1.eq}) is then used to determine the number of electrons $N_{A}^{1}$ of
each atom at the iteration $n=1$ 
\begin{equation}
N_{A}^{1}=Z_{A}-q_{A}^{1}=Z_{A}-\int d\mathbf{r~}\omega _{A}^{1}(\mathbf{r}%
)\rho (\mathbf{r}).
\end{equation}%
In the next iteration, the weight function $\omega _{A}^{2}(\mathbf{r})$ of
an atom $\left[ \text{Eq.\ (\ref{wn.eq})}\right] $ is constructed from its
atomic density $\rho _{A}^{1}(\mathbf{r})~$which integrates to $N_{A}^{1}=$ $%
\int d\mathbf{r}~\rho _{A}^{1}(\mathbf{r}).$~Because $N_{A}^{1}$ is
generally a non-integer number, $\rho _{A}^{1}(\mathbf{r})~$is built from a
linear interpolation between the densities of atoms with an integer number
of electrons smaller $(l)$ and larger $(u)$ than $N_{A}^{1}$: 
\begin{equation}
\rho _{A}^{1}(\mathbf{r})=(N_{A}^{1,u}-N_{A}^{1})\rho _{A}^{0,l}(\mathbf{r}%
)-(N_{A}^{1}-N_{A}^{1,lint})\rho _{A}^{0,u}(\mathbf{r}),
\end{equation}%
where $N_{A}^{1,u}=\int dr~\rho _{A}^{0,u}(\mathbf{r})$ and $%
N_{A}^{1,l}=\int dr~\rho _{A}^{0,l}(\mathbf{r})~$are the upper and lower
integer limits of $N_{A}^{1}$, respectively. The procedure is repeated until
the difference in atomic populations $N_{A}^{n}$ and $N_{A}^{n-1}$ of two
subsequent iterations $n$ and $n-1$ becomes zero. The self-consistent value
of the weight function is noted as $\omega _{A}(\mathbf{r})$ in the rest of
the paper.

The softness of an atom in the molecule is simply defined as: 
\begin{equation}
s_{A}\equiv \int d\mathbf{r}\ \omega _{A}(\mathbf{r})s(\mathbf{r})=\int d%
\mathbf{r}~s_{A}(\mathbf{r}),  \label{satomic.eq}
\end{equation}%
and the softness of a fragment $x$ is computed as the sum of the
contributions of the atoms belonging to the fragment: 
\begin{equation}
s_{x}=\sum_{A\ni x}s_{A}.  \label{sfragment.eq}
\end{equation}

The Hirshfeld-I method is applied to divide the local polarizability $\alpha
_{ij}(\mathbf{r})$ into condensed atomic polarizabilities. The dipole of an
atom $A$ in a molecule is defined as\cite{Hirshfeld.ref} 
\begin{equation}
\mathbf{P}^{A}\equiv -e\int d\mathbf{r~}\left[ \mathbf{r-R}^{A}\right] 
\mathbf{~}\rho _{A}(\mathbf{r}),  \label{dipAtom.eq}
\end{equation}%
The molecular dipole moment can be reconstructed exactly from atomic dipole
moments as follows 
\begin{equation}
\mathbf{P}=\sum_{A}\mathbf{P}^{A}+q_{A}e\mathbf{R}^{A}
\end{equation}
where $q_{A}\equiv Z_{A}-N_{A}$ is the atomic charge.

The \emph{intrinsic} condensed atomic polarizability of an atom $A$ is
defined by \textbf{the} derivative of the atomic dipole moment with respect
to the electric field: 
\begin{eqnarray}
\alpha _{ij}^{A} &=&\frac{\partial P_{i}^{A}}{\partial E_{j}},  \notag \\
&=&-e\int d\mathbf{r}\left[ r_{i}-R_{i}^{A}\right] \frac{\partial \rho _{A}(%
\mathbf{r})}{\partial E_{j}}.  \label{Halpha.eq}
\end{eqnarray}%
The total polarizability of the molecule is reconstructed by summing over
the atomic polarizabilities and adding the corresponding \emph{charge
delocalization contribution} 
\begin{equation}
\alpha _{ij}=\sum_{A}\left[ \alpha _{ij}^{A}+q_{A}^{(j)}R_{i}^{A}\right] ,
\label{alphatot.eq}
\end{equation}%
where the first-order perturbed atomic charge\textbf{\ }$q_{A}^{(j)}$\textbf{%
\ }is computed by using the relation\textbf{\ } 
\begin{equation}
q_{A}^{(j)}=e\frac{\partial q_{A}}{\partial E_{j}}=-e\int d\mathbf{r~}\frac{%
\partial \rho _{A}(\mathbf{r})}{\partial E_{j}}.  \label{qint.eq}
\end{equation}%
The total polarizability of a fragment $x$ is computed by summing over the
intrinsic polarizabilities of the atoms within the fragment and the \emph{%
intrafragmental }charge delocalization contribution as follows 
\begin{equation}
\alpha _{ij}^{x}\equiv \sum_{A\ni x}\left[ \alpha
_{ij}^{A}+q_{A}^{(j)}(R_{i}^{A}-\Omega _{i}^{x})\right] ,  \label{alphaSD.eq}
\end{equation}%
where $\mathbf{\Omega }_{{}}^{x}$ is the position of the center of mass of
fragment $x$. As shown in our previous studies\cite{PolarWater.ref}, the
definition of the charge delocalization contribution to the polarizability
of a fragment with respect to its center of mass ensures that identical
fragments at different positions in different molecules (such as the
backbone fragment in different amino acids or water molecules with identical
hydrogen bonds in different water clusters\cite{PolarWater.ref}) have
similar polarizability. In addition, by introducing the center of mass of
the fragment in the definition of its polarizability {Eq.\ (\ref{alphaSD.eq})%
}, we have shown that the remaining interfragmental charge delocalization
contribution to the polarizability of the molecule to which the fragment
belongs, i.e. $\sum_{x}\sum_{A\ni x}q_{A}^{(j)}(R_{i}^{A}-\Omega _{i}^{x})$,
is nearly constant for molecules of similar size.

Only the isotropic polarizabilities of fragments, defined as a third of the
trace of their polarizability tensor, will be discussed and presented.

\subsection{Softness dipole vector in the Hirshfeld scheme}

The contribution of the \textquotedblleft softness dipole
vector\textquotedblright\ to the polarizability of an atom in a molecule can
be computed as follows. The key quantity in Eqs.\ (\ref{Halpha.eq}) and (\ref%
{qint.eq}) is the derivative of the atomic electronic density relative to
the applied electric field : 
\begin{equation}
\frac{\partial \rho _{A}(\mathbf{r})}{\partial E_{j}}=w_{A}(\mathbf{r})\frac{%
\partial \rho (\mathbf{r})}{\partial E_{j}}.  \label{deriveAT.eq}
\end{equation}%
By using Eqs.\ (\ref{deriv.eq}) and (\ref{Berko-Parr.eq}), Eq.\ (\ref%
{deriveAT.eq}) can be written as follows 
\begin{eqnarray}
\frac{\partial \rho _{A}(\mathbf{r})}{\partial E_{j}} &=&w_{A}(\mathbf{r}%
)\int d\mathbf{r}^{\prime }\left[ \chi _{1}^{\mu }(\mathbf{r},\mathbf{r}%
^{\prime })+\frac{s(\mathbf{r})s(\mathbf{r}^{\prime })}{S}\right]
er_{j}^{\prime }  \notag \\
&=&\left[ \frac{\partial \rho _{A}(\mathbf{r})}{\partial Ej}\right] _{\mu
}+w_{A}(\mathbf{r})s(\mathbf{r})\left[ \frac{\int d\mathbf{r}^{\prime }s(%
\mathbf{r}^{\prime })er_{j}^{\prime }}{S}\right] ,  \notag \\
&=&\left[ \frac{\partial \rho _{A}(\mathbf{r})}{\partial Ej}\right] _{\mu
}+w_{A}(\mathbf{r})s(\mathbf{r})~\frac{D_{j}}{S},  \label{deriveAT2.eq}
\end{eqnarray}%
in which we have defined the derivative at a constant chemical potential
(first term of the righ-hand side of the last equality) and where $D_{j}~$is
the Cartesian component of the molecular softness dipole defined above $%
\left[ \text{Eq.\ (\ref{def-D.eq})}\right] $. Using Eqs.\ (\ref{Halpha.eq})
and (\ref{deriveAT2.eq}), the intrinsic condensed polarizability of an atom
can be expressed as follows 
\begin{eqnarray}
\alpha _{ij}^{A} &=&-\int d\mathbf{r}e\left[ r_{i}-R_{i}^{A}\right] \left[ 
\frac{\partial \rho _{A}(\mathbf{r})}{\partial Ej}\right] _{\mu }-\frac{D_{j}%
}{S}\int d\mathbf{r~}e\left[ r_{i}-R_{i}^{A}\right] w_{A}(\mathbf{r})s(%
\mathbf{r}),~  \notag \\
&=&\alpha _{ij}^{\mu ,A}-\frac{D_{j}}{S}\int d\mathbf{r~}e\left[
r_{i}-R_{i}^{A}\right] s_{A}(\mathbf{r}),  \notag \\
&=&\alpha _{ij}^{\mu ,A}-\frac{D_{j}D_{i}^{A}}{S},  \label{intdec.eq}
\end{eqnarray}%
where $s_{A}(\mathbf{r})$ has been defined above $\left[ \text{Eq.\ (\ref%
{satomic.eq})}\right] $ and the atomic softness dipole vector is defined as 
\begin{equation}
D_{i}^{A}\equiv e\int d\mathbf{r~}\left[ r_{i}-R_{i}^{A}\right] s_{A}(%
\mathbf{r}).  \label{def-DA.eq}
\end{equation}%
The \emph{intrinsic} polarizability of an atom can thus be decomposed into a
contribution at a constant chemical potential $\mu ~$and a dipole-softness
contribution 
\begin{equation}
\alpha _{ij}^{A}=\alpha _{ij}^{\mu ,A}+\alpha _{ij}^{A,DS},
\end{equation}%
with 
\begin{equation}
\alpha _{ij}^{A,DS}=-\frac{D_{i}^{A}D_{j}}{S}.  \label{alphaDS.eq}
\end{equation}

On the other hand, using Eqs.\ (\ref{qint.eq}) and (\ref{deriveAT2.eq}), the
charge induced by the electric field can be written as : 
\begin{equation}
q_{A}^{(j)}=-e\int d\mathbf{r}\frac{\partial \rho _{A}(\mathbf{r})}{\partial
E_{j}}=q_{A}^{\mu ,(j)}-\left( \frac{D_{j}s_{A}}{S}\right) .  \label{qdec.eq}
\end{equation}

The respective contributions to the total polarizability of the molecule are
again reconstructed by adding the atomic contributions and the respective
charge delocalization contributions: 
\begin{equation}
\alpha _{ij}^{\mu }=\sum_{A}\left[ \alpha _{ij}^{\mu ,A}+q_{A}^{\mu
,(j)}eR_{i}^{A}\right]   \label{alpha_mu_tot.eq}
\end{equation}%
and 
\begin{equation}
\alpha _{ij}^{DS}=\sum_{A}\left[ \alpha _{ij}^{A,DS}-\left( \frac{D_{j}s_{A}%
}{S}\right) eR_{i}^{A}\right]   \label{alpha_ds_tot.eq}
\end{equation}%
Eq.\ (\ref{alpha_ds_tot.eq}) allows us to compute the contribution of the
dipole softness polarizability to the polarizability of a fragment [Eq.\ (%
\ref{alphaSD.eq})] by summing in Eq.\ (\ref{alpha_ds_tot.eq}) over the atoms
of a fragment: 
\begin{equation}
\alpha _{ij}^{x,DS}=\sum_{A\ni x}\left[ \alpha _{ij}^{A,DS}-\left( \frac{%
D_{j}s_{A}}{S}\right) eR_{i}^{A}\right] .  \label{alpha_x_ds.eq}
\end{equation}%
It should be emphasized that in the definition of the dipole-softness
polarizability [Eq.\ (\ref{alpha_x_ds.eq})] we do not compute the charge
delocalization relative to the center of mass of the fragment as it was done
in the definition of the polarizability of a fragment [Eq.\ (\ref{alphaSD.eq}%
)]. Indeed, the molecular dipole softness $D_{j}$ is already defined
relative to the center of mass of the molecule. On the other hand, there is
no reason to enforce the transferability of the dipole softness contribution
to the polarizability. Indeed, the contribution of a fragment $x$ to the
molecular dipole softness, $D_{j}^{x}\equiv \int d\mathbf{r~}es_{x}~\mathbf{r%
}$, is expected to be different for identical fragments in different
molecules on the opposite to $\alpha _{ij}^{x}$. As will be discussed below,
the local softness $s_{1}$ $=$ $S-s_{2~}$of a backbone (which is different
from $\overline{S}_{1}$ in Eq.\ (\ref{eq:linear}), see Eq.\ (23) in Ref.\ (%
\ref{CoulombHole.ref}) for more details) is not identical in two different
amino acids but depends strongly on the relative reactivity of the rest of
the molecule, namely the side chain; the softness $s_{1}~$of the backbone
will be large in amino acids with a \textquotedblleft
hard\textquotedblright\ side chain ($s_{2}$ small as in glycine) and small
in amino acids with a \textquotedblleft soft\textquotedblright\ side chain ($%
s_{2}$ large as in phenylalanine). For this reason one can expect the dipole
softness polarizability to be a non-transferable property.

The contribution of dipole-softness to the polarizability can also be
analyzed in an alternative way. The softness dipole vector of a molecule is
easily decomposed into contributions of fragments in the Hirshfeld scheme.
These fragment softness dipole vectors are related to the atomic softness
dipole vectors [Eq.\ (\ref{def-DA.eq}] as follows 
\begin{eqnarray}
D_{x} &=&\int d\mathbf{r}s_{x}(\mathbf{r})e\mathbf{r},  \notag \\
&=&\sum_{A\ni x}\int d\mathbf{r}s_{A}(\mathbf{r})e\mathbf{r},  \notag \\
&=&\sum_{A\ni x}\int d\mathbf{r}s_{A}(\mathbf{r})e(\mathbf{r}-\mathbf{R}^{A}+%
\mathbf{R}^{A}),  \notag \\
&=&\sum_{A\ni x}\mathbf{D^{A}}+s_{A}{R^{A}.}
\end{eqnarray}%
Using the definition Eq.\ (\ref{def-D.eq}), one finds 
\begin{eqnarray}
\mathbf{D} &=&\int d\mathbf{r\ }s_{1}(\mathbf{r})e\mathbf{r+}\int d\mathbf{%
r\ }s_{2}(\mathbf{r})e\mathbf{r},  \notag \\
&=&\mathbf{D}_{1}+\mathbf{D}_{2},  \label{vector-D.eq}
\end{eqnarray}%
where $s_{1}$ and $s_{2}$ are computed by using Eq.\ (\ref{sfragment.eq}).
Therefore the dipole softness contribution $\frac{-D^{2}}{3S}$to the
polarizability of a molecule [Eq.(\ref{pol-soft-trace.eq})] can be written
as the sum of three terms 
\begin{eqnarray}
\frac{-D^{2}}{3S} &=&\alpha _{iso}-\alpha _{iso}^{\mu },  \notag \\
&=&-\frac{D_{1}^{2}}{3S}-\frac{D_{2}^{2}}{3S}-\frac{2}{3S}\mathbf{D}_{1}.%
\mathbf{D}_{2}.  \label{vector-D-pol.eq}
\end{eqnarray}%
This last equation allows to separate the dipole-softness contribution to
the polarizability into contributions of each fragment and a coupling term
between the two fragments.

\subsection{Computational method}

The geometries of the twenty amino acids in their neutral form were
calculated during a previous study using the MP2/6-311G(d,p) basis set in
Gaussian03.\cite{Gaussian} The global polarizabilities were calculated using
the same level of theory and basis set. The global softness was obtained by
using the energies of the frontier molecular orbitals of the SCF density 
\begin{equation}
S=\frac{1}{(\epsilon _{LUMO}-\epsilon _{HOMO})}.  \label{SP.eq}
\end{equation}%
The dipole softness of the amino acids was computed using the Brabo package.%
\cite{Brabo} The softnesses, polarizabilities and dipole softness of the
backbone and side chain were evaluated using the Stock program\cite%
{Rousseau.ref}, part of the Brabo package.\cite{Brabo} The MP2 density was
used for the evaluation of the Hirshfeld-I weight function~$\omega _{A}(%
\mathbf{r})$. The local softness $s_{A}$ of each atom in Eq.\ (\ref%
{satomic.eq}) was computed using the frontier-orbital approximation: . 
\begin{equation}
s_{A}\approx \frac{\left[ \int d\mathbf{r}\omega _{A}(\mathbf{r})\rho
_{HOMO}(\mathbf{r})+\int d\mathbf{r}\omega _{A}(\mathbf{r})\rho _{LUMO}(%
\mathbf{r})\right] }{2\left[ \epsilon _{LUMO}-\epsilon _{HOMO}\right] },
\label{fpA.eq}
\end{equation}%
where the HOMO and LUMO densities are also extracted from the SCF density.
The global softness of the common backbone fragment and the Coulomb hole $%
q_{g}^{h}$ of the amino acids family were calculated from Eq.\ (\ref%
{eq:linear}) by applying a linear regression to the $S/s_{2}$ curve.

In the frontier-orbital approximation the softness dipole vector is computed
using the Brabo\cite{Brabo} program as 
\begin{equation}
D_{i}=\frac{S}{2}\int d\mathbf{r}\left[ \rho _{HOMO}(\mathbf{r})+\rho
_{LUMO}(\mathbf{r})\right] r_{i}e.  \label{HLdip.eq}
\end{equation}

\section{Results and Discussion}

The list of the twenty amino acids with their names, side chains and
reactivity groups is given in Table 1. The amino acids are usually divided
into eight different (reactivity) groups according to the nature of the side
chain.\cite{Proteins.ref}

Global softness ($S$), local softness of the backbone ($s_{1}$) and side
chain ($s_{2}$), global polarizability ($\alpha $) and polarizability of the
backbone ($\alpha_{1}$) and side chain ($\alpha _{2} $) of these amino acids
are presented in Table 2. Figure 1 depicts the global softness as function
of the local softness of the side chain. The linear correlation between
these two quantities amounts to $R=0.9242$, while the curve also
demonstrates a clear separation of the amino acids into seven reactivity
groups recovering to a large extent those defined in Table 1. It must be
noted that the hydroxylic amino acids, as well as the proline amino acid,
are grouped together with the aliphatic amino acids in the rest of the
discussion of the results as the values of their softnesses are overlapping.
The values of the local softness $s_{2}$ of the side chains follow the
following order: alphatic$<$acidic$<$amide$<$basic$<$sulphur-containing$<$%
histidine$<$aromatic. At first glance, this order agrees with the chemical
intuition. Indeed, one expects the aromatic amino acids, the most
polarizable members of the family, to be the \textquotedblleft
softest\textquotedblright . On the opposite, the aliphatic amino acids, the
less polarizable molecules of the family, are expected to be the
\textquotedblleft hardest\textquotedblright . This reasoning is based on the
empirical rule that a polarizable atom is also a soft atom. One will see
below that \textquotedblleft this rule of thumb\textquotedblright\ must be
used however with care.

Fig.\ 1 can be compared to Fig.\ 4 in Ref.\ \ref{CoulombHole.ref}, where the
global softness of the amino acids, calculated using the same method and at
identical geometries, is depicted as a function of the local softness of the
side chain calculated using the method of Contreras et al.\cite%
{Contreras.ref} Although the values of the local softness of the side chains
differ from the values listed in Table 2 in Ref.\ \ref{CoulombHole.ref}, the
group separation is completely reproduced using the Hirshfeld-I method,
demonstrating the robustness of the softness as a reactivity descriptor and
its independence on the details of the partitioning method.

The Coulomb hole and global (average) softness of the backbone fragment $%
\overline{S}_{1}$ [Eq.\ (\ref{eq:linear})], derived by applying a linear
regression to the curve $S(s_{2})$ depicted in Fig.\ 1, are 0.7439 and
1.7113, respectively. Those values differ by only 0.5\% from the values
listed in Table 1 of Ref.\ \ref{CoulombHole.ref}, being 0.7476 and 1.7196,
respectively. The concept of the \textquotedblleft Coulomb
hole\textquotedblright\ is thus further confirmed, while its independence on
the partitioning method is clearly demonstrated. \ 

On the other hand, the partitioning of amino acids into subgroups observed
in Figure 1 and in Ref.\ (\ref{CoulombHole.ref}) can be explained as
follows. The global softness $S$ in Table 2 shows relatively little
variation among the twenty amino acids, being the smallest for aspartic acid
(1.78 au) and largest for tryptophan (2.54 au). On the opposite, the
partitioning of the global softness $S$ between the fragments [Eq.\ (\ref%
{sfragment.eq})] varies considerably. For instance, the local softness of
the side chain $s_{2}$ varies between 0.10 au (glycine) and 2.45
(tryptophan). A large similar variation is obviously observed for the
softness of the backbone $s_{1}$ as by definition $s_{1}=S-s_{2}$. As shown
in Figure 1, points in the (S,s)$_{2}$ map which are close identify residues
belonging to the same sub-family. The results of Figure 1 can be summarized
by saying that \textquotedblleft similar molecules have similar global
softness and similar contributions of their corresponding fragments to the
global softness\textquotedblright .

Figure 2 shows the strong correlation ($R=0.9975$) between the global
molecular polarizability of an isolated amino acid and the polarizability of
its side chain. Therefore the polarizability of an amino acid is largely
determined by the polarizability of its side chain. On the other hand, the
value of the polarizability of the backbone is similar for all isolated
amino acids and has a value close to its average value (28.55$\pm $0.81 au),
except for proline (24.75 au) for which the side chain is bounded to the
backbone on the contrary to the other amino acids.

Another remarkable feature of the polarizability of an isolated amino acid
is a strong correlation ($R=0.9659$) between its polarizability and its
number of electrons $N$ as can be seen in Figure 3. This correlation is a
consequence of the relation between the polarizability and the molecular
volume\cite{Politzer2.ref}. Indeed, molecular polarizability and molecular
volume can be related to each other by using simple electrostatic models.
For instance, for a spherical molecule of radius $R$ represented by a
dielectric sphere with a macroscopic dielectric constant $\varepsilon ,$ the
polarizability $\alpha _{sphere}~$is given by \cite{Lambin.ref} 
\begin{eqnarray}
\alpha _{sphere} &=&\frac{R^{3}}{4\pi \varepsilon _{0}}\left( \frac{%
\varepsilon -1}{\varepsilon +1}\right) ,  \label{CM.eq} \\
&=&\frac{V}{16\pi ^{2}\varepsilon _{0}}\left( \frac{\varepsilon -1}{%
\varepsilon +1}\right) ,
\end{eqnarray}%
where $\varepsilon _{0}~$is the vacuum permitivity and $V$ the molecular
volume. Assuming a homogeneous molecular electronic density, 
\begin{equation}
\rho =\frac{N}{V},
\end{equation}%
one finds 
\begin{equation}
\frac{\alpha _{sphere}}{N}=\frac{1}{16\pi ^{2}\varepsilon _{0}\rho }\left( 
\frac{\varepsilon -1}{\varepsilon +1}\right) .  \label{alpha-per-N.eq}
\end{equation}%
In this model, the polarizability per electron would be constant for
different molecules of different sizes (measured in the model by the
parameter $R$) but having similar properties (controled in the model by the
values of the average density $\rho $ and dielectric constant $\varepsilon $%
). In the case of the amino acids, it must be noted that the approximated
linear relation between the molecular size and the molecular polarizability
is a consequence of the similarity of the examined molecules, which allows
to describe them as a molecular family with $\alpha /N$ varying between 0.85
au (Asp) to 1.2 au (Trp).

Softness and polarizability have always been intuitively related to each
other as mentioned above. To test this relation, we have reproduced in
Figure 4 the global polarizability per electron global $\alpha /N$ of the
amino acids in function of their softness $S$. One observes some qualitative
features, for instance, as mentioned above, the aromatic amino acids are the
softest and the most polarizable. The correlation coefficient of the linear
regression represented in Figure 4 is only $0.7297$. Therefore, one cannot
conclude that softness and polarizability are strictly proportional. An
example is for instance the couple serine and isoleucine, which have nearly
the same softness $S$ $\approx 1.82$ au but different polarizabilities $%
\alpha /N\approx 0.85$\textbf{\ }$[\alpha \approx 48.21]~$au for Ser and $%
\alpha /N\approx 1.1$ $[\alpha \approx 78.65]$ au for Ile.

As was shown in the method section, the lack of proportionality between
polarizability and softness is not surprizing since polarizability does not
depend directly on the softness but also on the softness dipole vector
through the term $\frac{-D^{2}}{3S}$ in Eq.\ (\ref{pol-soft-trace.eq}).
Table 3 lists the $\alpha^{DS}$ contributions to the polarizabilities of the
amino acids, their backbones ($\alpha^{DS}_{1}$) and their side chains ($%
\alpha^{DS}_{2}$), calculated using Eq.\ \ref{alpha_x_ds.eq}.

Because the partitioning of the global softness $S$ between the fragments
(backbone plus side chain) varies considerably among the amino acids (Table
2), one expects that the dipole-softness polarizability to vary
significantly. The dipole-softness polarizability $\alpha ^{DS}$ ranges from
-0.004 au for aspartic acid to -27.78 au for histidine. The value for
histidine is also significantly larger than the rest of the values, the
aromatic amino acids having values around -10 au. Also the distribution of
the dipole-softness polarizability between the backbone ($\alpha _{1}^{DS}$%
)~and the side chain ($\alpha _{2}^{DS}$)~differs from the distribution of
the corresponding local softnesses. One observes indeed that local softness
values (Table 2) of the side chains ($s_{2}$) increase in the order alphatic$%
<$acidic$<$amide$<$basic$<$sulphur-containing$<$histidine$<$aromatic, while
the values for the backbone ($s_{1}$)~decrease in the same order. On the
opposite, the variation of the values of the dipole-softness polarizability
of the side chains $\alpha _{2}^{DS}$ or of the backbone $\alpha _{1}^{DS}$
of the amino acids (Table 3) does not follow the classification of the
residues in the biochemical families. In other words, amino acids with
similar $\alpha _{2}^{DS}~$ can be in different families as for instance Ala 
$\alpha _{2}^{DS}=-0.67~$ au and Met $\alpha _{2}^{DS}=-0.62$ au. However
one observes that absolute large values $|\alpha ^{DS}|> 2.7$ for aromatic,
amide, basic and sulfur-containing amino acids correspond also to large
contributions of the side chains $|\alpha ^{DS}| \simeq |\alpha _{2}^{DS}|$.

Figure 5 shows the very good correlation between the dipole-softness
polarizability values of the amino acid and those of the side chain, with a
correlation coefficient of $R=0.9890$. This plot also allows some separation
of the amino acids into reactivity groups, similar to Figure 1, although the
order is different showing more overlap. The amino acids are divided into
three major groups, the first one containing the acidic, aliphatic and amide
amino acids, the second one containing the sulphur-containing, basic and
aromatic amino acids and the third group contains histidine, which has a
remarkably larger dipole-softness contribution than the rest.

Table 4 contains the dipole-softness contributions partitioned using Eq.\ (%
\ref{vector-D-pol.eq}), where the total contribution to the molecular
polarizability $-\frac{D^{2}}{3S}$ is shown (identical to the $\alpha ^{DS}$
in Table 3) together with the contribution of the backbone $-\frac{D_{1}^{2}%
}{3S}$, the side chain $-\frac{D_{2}^{2}}{3S}$ and the interfragmental
contribution $-\frac{2\mathbf{D}_{1}\mathbf{D_{2}}}{3S}$. In this
partitioning, the dipole-softness polarizabilities of the fragments are
negative for the backbone and the side chain, whereas the interfragmental
contribution is positive for most amino acids excep Gly and Ala. Remarkably,
despite the difference in the two approaches [Eqs.\ (\ref{alpha_x_ds.eq})
and (\ref{vector-D-pol.eq})], the values for the side chains listed in
Tables 3 and 4 correlate considerably, with a correlation coefficient of $%
R=0.9905$. Also the group separation observed in Figure 6 by plotting the
dipole-softness polarizabilities as function of $-\frac{D_{2}^{2}}{3S}$, is
similar to the one seen in Figure 5, with a slightly lower correlation
coefficient of $R=0.9637$. However, the contributions of the backbones,
which do not become more negative than -3.2 au in Table 4, do not correlate
with the dipole-softness polarizability values of the backbones reported in
Table 3. The interfragmental contribution does not reflect the biochemical
classification.

One can conclude from Tables 3 and 4 that dipole-softness polarizability is
a descriptor which permits to group the amino acids by similarity of their
dipole-softness polarizabilities but this classification does not reflect
any biochemical classification.

\section{Conclusions}

The molecular family of amino acids was studied by defining and computing
the local softness and the local polarizability of their side chains. The
concepts of the local softness of a fragment and of the Coulomb hole of a
molecular family, that were introduced in Ref.\ \ref{CoulombHole.ref}, were
found independent of the partitioning method of the electronic properties:
both the Contreras et al.\ method (Ref. \ref{CoulombHole.ref}) and the
Hirshfeld-I method (present work) gave similar results. One concludes that
the softness of the side chains allowes to separate the amino acids into
groups reflecting their usual biochemical classification. Amino acids within
of the same biochemical family are close to each other in the map (S, s$_{%
\mathbf{2}}$) (Fig.\ 1) which means that similar molecules have both similar
global and local softnesses.

The polarizability of the twenty amino acids and of their side chains were
also computed. The global polarizability was found approximately
proportional to the number of electrons $N$ of the amino acid. Indeed, $%
\alpha /N$ varies between 0.85 au (Asp) to 1.2 au (Trp) within the whole set
of amino acids. These results reflect the well-known proportionality between
molecular polarizability and molecular size or volume [See Eq.\ (\ref%
{volume.eq})].\cite{Politzer2.ref} \ Consequently, the polarizability of the
side chains does not reflect the biochemical classification of the amino
acids on the contrary of their local softness.

The statement that systems which are more polarizable than other are also
softer was tested by comparing the softness and polarizability of the side
chains of the amino acids. A very moderate correlation was found to exist
between global softness and global polarizability per electron. It is shown
that the global softness $S$ of two amino acids may be identical whereas
their polarizability $\alpha $ differ significantly.

Finaly, a relation was derived between the local polarizability of a
molecule and its local softness by applying the Berkowitch-Parr relation.
The key quantity is the \textquotedblleft softness dipole
vector\textquotedblright\ of a fragment in Eq.\ (\ref{def-D.eq}). We found
that the dipole-softness contribution to polarizability exhibits very
different behavior than the local softness, explaining the lack of
proportionality between softness and polarizability. The softness dipole
vector contribution to the polarizability, as well as the polarizability
itself, does not reflect the biochemical classification of the amino acids:
for example, acidic and amide amino acids have as low values of
dipole-softness contribution to the their polarizability as aliphatic amino
acids, while histidine has a value of the dipole-softness contribution to
the its polarizability almost three times as large as the value for aromatic
amino acids.

From the results presented in this paper one can conclude that the often
assumed relation between softness and polarizability is not as
straightforward as one might think, and that although polarizability and
dipole softness undoubtedly reflect somehow the reactivity of a molecule,
their values are not good descriptors of the molecular similarity, in
contrast to softnesses.

\section{Acknowledgments}

A.K.\ and C.V.A.\ acknowledge the Flemish FWO for research grant nr.\
G.0629.06. We gratefully acknowledge the University of Antwerp for the
access to the university's CalcUA supercomputer cluster.

\renewcommand{\baselinestretch}{1} \newpage

\section{Tables}

\begin{table}[th]
\centering
\begin{tabular}{llll}
\hline
Name & Abbrevation & Side Chain & Group \\ \hline
Aspartic Acid & Asp & \textsc{-ch$_{2}$cooh} & Acidic \\ 
Glutamic Acid & Glu & \textsc{-ch$_{2}$ch$_{2}$cooh} & Acidic \\ 
Alanine & Ala & \textsc{-ch$_{3}$} & Aliphatic \\ 
Glycine & Gly & \textsc{-h} & Aliphatic \\ 
Isoleucine & Ile & \textsc{ch(ch$_{3}$)ch$_{2}$ch$_{3}$} & Aliphatic \\ 
Leucine & Leu & \textsc{-ch$_{2}$ch(ch$_{3}$)$_{2}$} & Aliphatic \\ 
Proline & Pro & \textsc{-c$_{3}$h$_{6}$(ring)} & Aliphatic \\ 
Serine & Ser & \textsc{-ch$_{2}$oh} & Hydroxylic \\ 
Threonine & Thr & \textsc{-ch(oh)ch$_{3}$} & Hydroxylic \\ 
Valine & Val & \textsc{-ch(ch$_{3}$)$_{2}$} & Aliphatic \\ 
Asparagine & Asn & \textsc{-ch$_{2}$conh$_{2}$} & Amide \\ 
Glutamine & Gln & \textsc{-ch$_{2}$ch$_{2}$conh$_{2}$} & Amide \\ 
Phenylalanine & Phe & \textsc{-ch$_{2}$(phenyl)} & Aromatic \\ 
Tryptophan & Trp & \textsc{-ch$_{2}$(c$_{2}$h$_{2}$n)(phenyl)} & Aromatic \\ 
Tyrosine & Tyr & \textsc{-ch$_{2}$(phenyl)oh} & Aromatic \\ 
Arginine & Arg & \textsc{-ch$_{2}$ch$_{2}$ch$_{2}$nhc(nh)nh$_{2}$} & Basic
\\ 
Lysine & Lys & \textsc{-ch$_{2}$ch$_{2}$ch$_{2}$ch$_{2}$nh$_{2}$} & Basic \\ 
Histidine & His & \textsc{-ch$_{2}$c$_{3}$n$_{2}$h$_{3}$(ring)} & Histidine
\\ 
Cysteine & Cys & \textsc{-ch$_{2}$sh} & Sulphur-containing \\ 
Methionine & Met & \textsc{-ch$_{2}$ch$_{2}$sch$_{3}$} & Sulphur-containing
\\ \hline
\end{tabular}%
\caption{The names, abbrevations, and formula's of the side chain and of the
reactivity groups of the twenty amino acids examined in this study.}
\end{table}
\newpage 
\begin{table}[th]
\centering
\begin{tabular}{lcccccc}
\hline
Name & $S$ & s$_{1}$ & s$_{2}$ & $\alpha$ & $\alpha_{1}$ & $\alpha_{2}$ \\ 
\hline
Asp & 1.78 & 1.12 & 0.66 & 59.76 & 27.99 & 21.14 \\ 
Glu & 1.88 & 1.37 & 0.51 & 70.38 & 28.48 & 31.69 \\ 
Ala & 1.85 & 1.56 & 0.29 & 45.52 & 29.83 & 9.55 \\ 
Gly & 1.84 & 1.74 & 0.10 & 34.45 & 30.46 & 1.09 \\ 
Ile & 1.82 & 1.40 & 0.42 & 78.65 & 27.88 & 38.29 \\ 
Leu & 1.85 & 1.38 & 0.46 & 78.90 & 28.58 & 38.82 \\ 
Pro & 1.88 & 1.33 & 0.56 & 63.26 & 24.75 & 27.41 \\ 
Ser & 1.82 & 1.48 & 0.34 & 48.21 & 29.16 & 11.76 \\ 
Thr & 1.84 & 1.34 & 0.50 & 59.31 & 27.77 & 22.41 \\ 
Val & 1.81 & 1.42 & 0.39 & 67.57 & 27.84 & 29.20 \\ 
Asn & 1.84 & 0.94 & 0.90 & 64.08 & 27.93 & 25.21 \\ 
Gln & 1.87 & 1.15 & 0.72 & 75.27 & 27.99 & 34.26 \\ 
Phe & 2.22 & 0.14 & 2.08 & 104.59 & 28.00 & 61.31 \\ 
Trp & 2.53 & 0.08 & 2.45 & 132.92 & 27.76 & 89.30 \\ 
Tyr & 2.34 & 0.12 & 2.23 & 109.75 & 28.09 & 65.17 \\ 
Arg & 2.12 & 0.71 & 1.41 & 100.14 & 27.74 & 54.89 \\ 
Lys & 1.91 & 0.67 & 1.24 & 86.51 & 28.72 & 42.46 \\ 
His & 2.19 & 0.04 & 2.15 & 84.83 & 27.35 & 42.55 \\ 
Cys & 2.00 & 0.35 & 1.66 & 62.03 & 29.10 & 23.78 \\ 
Met & 2.10 & 0.58 & 1.52 & 83.96 & 29.15 & 44.78 \\ \hline
\end{tabular}%
\caption{Global softness $S$ (Eq.\ (\protect\ref{SP.eq})), local softness of
the backbone and side chain $s_{1}$ and $s_{2}$ (Eq.\ (\protect\ref%
{sfragment.eq})), calculated from the SCF density using the 6-311G(d,p)
basis set, and global polarizability $\protect\alpha $ and local
polarizabilities of the backbone and side chain $\protect\alpha _{1}$ and $%
\protect\alpha _{2} $ (Eq.\ (\protect\ref{alphaSD.eq})) of the twenty amino
acids, calculated at the MP2/6-311G(d,p) level of theory. All quantities are
in au. }
\end{table}
\newpage 
\begin{table}[th]
\centering
\begin{tabular}{lccc}
\hline
Name & $\alpha^{DS}$ & $\alpha^{DS}_{1}$ & $\alpha^{DS}_{2}$ \\ \hline
Asp & -0.04 & -0.15 & 0.11 \\
Glu & -1.76 & -2.05 & 0.30 \\ 
Ala & -0.98 & -0.67 & -0.31 \\ 
Gly & -1.65 & -1.50 & -0.16 \\ 
Ile & -2.69 & -2.67 & -0.02 \\ 
Leu & -2.09 & -2.24 & 0.15 \\ 
Pro & -1.47 & -0.86 & -0.61 \\ 
Ser & -2.12 & -1.88 & -0.25 \\ 
Thr & -1.51 & -1.19 & -0.32 \\ 
Val & -1.99 & -1.79 & -0.21 \\ 
Asn & -1.46 & 0.59 & -2.05 \\ 
Gln & -1.14 & -1.38 & 0.24 \\ 
Phe & -11.18 & 0.48 & -11.66 \\ 
Trp & -9.82 & 0.23 & -10.05 \\ 
Tyr & -9.19 & 0.35 & -9.54 \\ 
Arg & -6.07 & 1.22 & -7.28 \\ 
Lys & -7.89 & 1.25 & -9.14 \\ 
His & -27.78 & 0.17 & -27.95 \\ 
Cys & -8.91 & 0.04 & -8.96 \\ 
Met & -5.94 & -0.62 & -5.31 \\ \hline
\end{tabular}%
\caption{Global dipole-softness polarizability $\protect\alpha^{DS}$, local
dipole-softness polarizabilities of the backbone $\protect\alpha^{DS}_{1}$
and side chain $\protect\alpha^{DS}_{2}$ of the twenty amino acids,
calculated from the SCF density obtained with the 6-311G(d,p) basis set,
using Eq.\ (\protect\ref{alpha_x_ds.eq}). All quantities are in au.}
\end{table}
\newpage 
\begin{table}[th]
\centering
\begin{tabular}{lcccc}
\hline
Name & $-\frac{D^{2}}{3S}$ & $-\frac{D^{2}_{1}}{3S}$ & $-\frac{D^{2}_{2}}{3S}
$ & $-\frac{2\mathbf{D}_{1}\mathbf{D}_{2}}{3S}$ \\ \hline
Asp & -1.46 & -1.68 & -1.43 & 3.07 \\ 
Glu & -1.76 & -3.03 & -0.68 & 1.95 \\ 
Ala & -0.99 & -0.63 & -0.27 & -0.08 \\ 
Gly & -1.66 & -1.38 & -0.04 & -0.23 \\ 
Ile & -2.69 & -3.23 & -0.59 & 1.13 \\ 
Leu & -2.09 & -2.97 & -0.58 & 1.46 \\ 
Pro & -1.47 & -1.06 & -0.82 & 0.40 \\ 
Ser & -2.13 & -1.91 & -0.29 & 0.07 \\ 
Thr & -1.51 & -1.44 & -0.57 & 0.49 \\ 
Val & -1.99 & -2.19 & -0.61 & 0.81 \\ 
Asn & -0.04 & -1.00 & -3.65 & 3.19 \\ 
Gln & -1.14 & -2.87 & -1.26 & 2.98 \\ 
Phe & -11.18 & -0.02 & -12.16 & 1.00 \\ 
Trp & -9.82 & -0.01 & -10.28 & 0.47 \\ 
Tyr & -9.19 & -0.02 & -9.91 & 0.73 \\ 
Arg & -6.07 & -2.36 & -10.86 & 7.16 \\ 
Lys & -7.89 & -1.49 & -11.88 & 5.49 \\ 
His & -27.78 & 0.00 & -28.12 & 0.34 \\ 
Cys & -8.91 & -0.09 & -9.09 & 0.27 \\ 
Met & -5.94 & -0.92 & -5.61 & 0.60 \\ \hline
\end{tabular}%
\caption{Global dipole-softness polarizability $-\frac{D^{2}}{3S}$,
dipole-softness polarizabilities of the backbone $-\frac{D^{2}_{1}}{3S}$,
side chain $-\frac{D^{2}_{2}}{3S}$ and the interfragmental contribution $-%
\frac{2\mathbf{D}_{1}\mathbf{D}_{2}}{3S}$ of the twenty amino acids,
calculated from the SCF density obtained with the 6-311G(d,p) basis set
using Eq.\ (\protect\ref{vector-D-pol.eq}). All quantities are in au.}
\end{table}
\newpage

\section{Captions}

\noindent Figure 1: The global softness $S$ of the amino acids as function
of the local softness $s_{2}$ of the side-chain, calculated from the SCF
density using the 6-311G(d,p) basis set. All values are depicted in au.

\noindent Figure 2: The global polarizability $\alpha $ of the amino acids
as function of the local polarizability $\alpha _{2}$ of the side-chain 
[Eq.\ (\ref{alphaSD.eq})], calculated at the MP2/6-311G(d,p) level.
All values are depicted in au.

\noindent Figure 3: The linear dependence of the polarizabilities of amino
acids $\alpha $, calculated at the MP2/6-311G(d,p) level, on the number of
electrons $N$.

\noindent Figure 4: The polarizability per electron $\alpha /N$, calculated
at the MP2/ 6-311G(d,p) level, as function of the global softness $S$ of the
amino acids, calculated from the SCF density obtained with the 6-311G(d,p)
basis set. All values are depicted in au.

\noindent Figure 5: The global dipole-softness polarizability $-\alpha ^{DS}$
as function of the the local dipole-softness polarizabilities of the side
chain $-\alpha _{2}^{DS}$, calculated from the SCF density obtained with the
6-311G(d,p) basis set. All values are depicted in au.

\noindent Figure 6: The global dipole-softness polarizability $\frac{D^{2}}{%
3S}$ as function of the the local dipole-softness polarizabilities of the
side chain $\frac{D_{2}^{2}}{3S}$, calculated from the SCF density obtained
with the 6-311G(d,p) basis set. All values are depicted in au. \newpage

\section{Figures}

\begin{figure}[th]
\resizebox{0.9\textwidth}{!}{\includegraphics{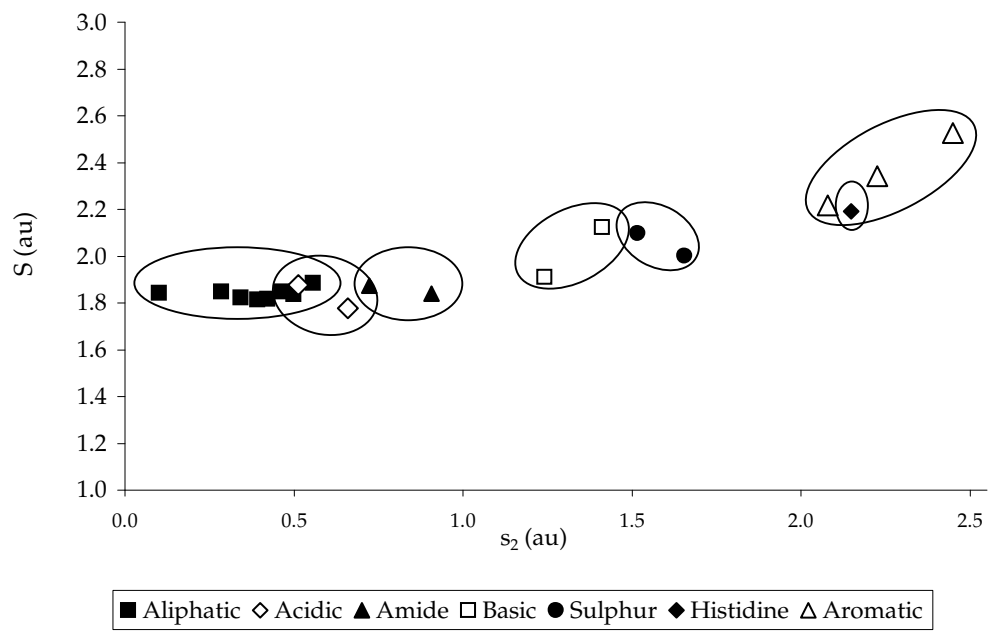}}
\caption{The global softness $S$ of the amino acids as function of the local
softness $s_{2}$ of the side-chain, calculated from the SCF density using
the 6-311G(d,p) basis set. All values are depicted in au.}
\end{figure}
\newpage 
\begin{figure}[th]
\resizebox{0.9\textwidth}{!}{\includegraphics{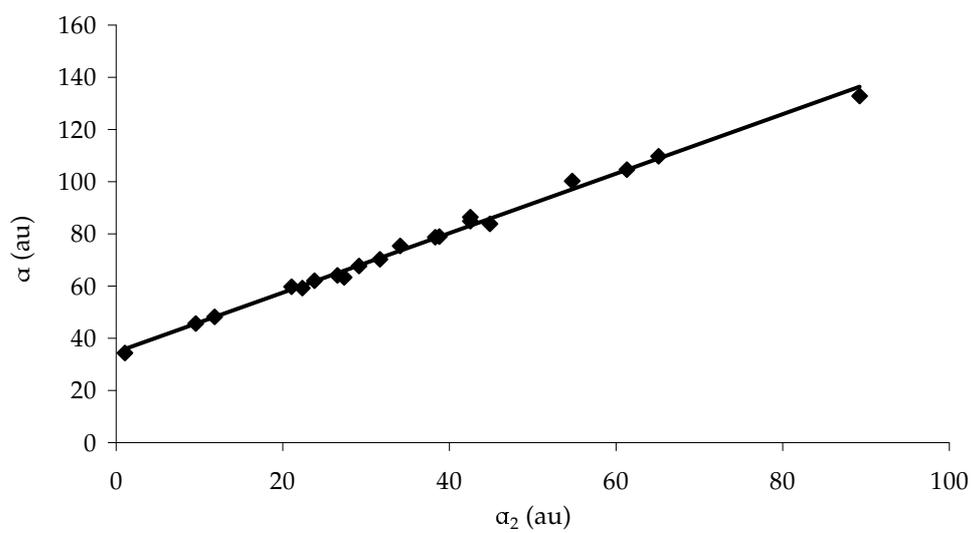}}
\caption{The global polarizability $\protect\alpha$ of the amino acids as
function of the local polarizability $\protect\alpha_{2}$ of the side-chain [Eq.\ (\ref{alphaSD.eq})],
calculated at the MP2/6-311G(d,p) level. All values are depicted in au.}
\end{figure}
\newpage 
\begin{figure}[th]
\resizebox{0.9\textwidth}{!}{\includegraphics{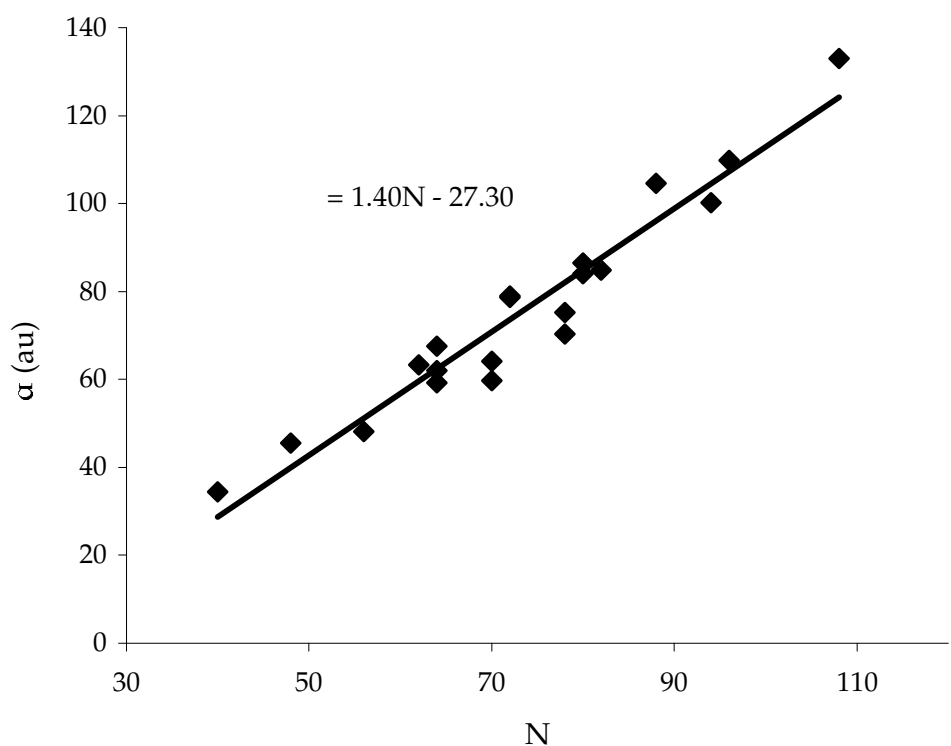}}
\caption{The linear dependence of the polarizabilities of amino acids $%
\protect\alpha$, calculated at the MP2/6-311G(d,p) level, on the number of
electrons $N$.}
\end{figure}
\newpage 
\begin{figure}[th]
\resizebox{\textwidth}{!}{\includegraphics{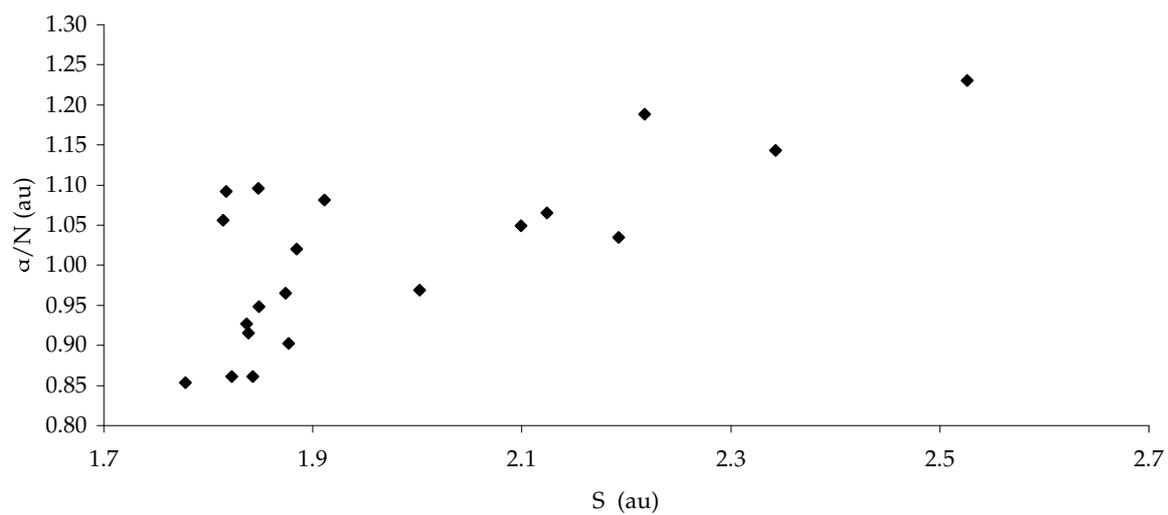}}
\caption{The polarizability per electron $\protect\alpha /N$, calculated at
the MP2/ 6-311G(d,p) level, as function of the global softness $S$ of the
amino acids, calculated from the SCF density obtained with the 6-311G(d,p)
basis set. All values are depicted in au.}
\end{figure}
\newpage 
\begin{figure}[th]
\resizebox{\textwidth}{!}{\includegraphics{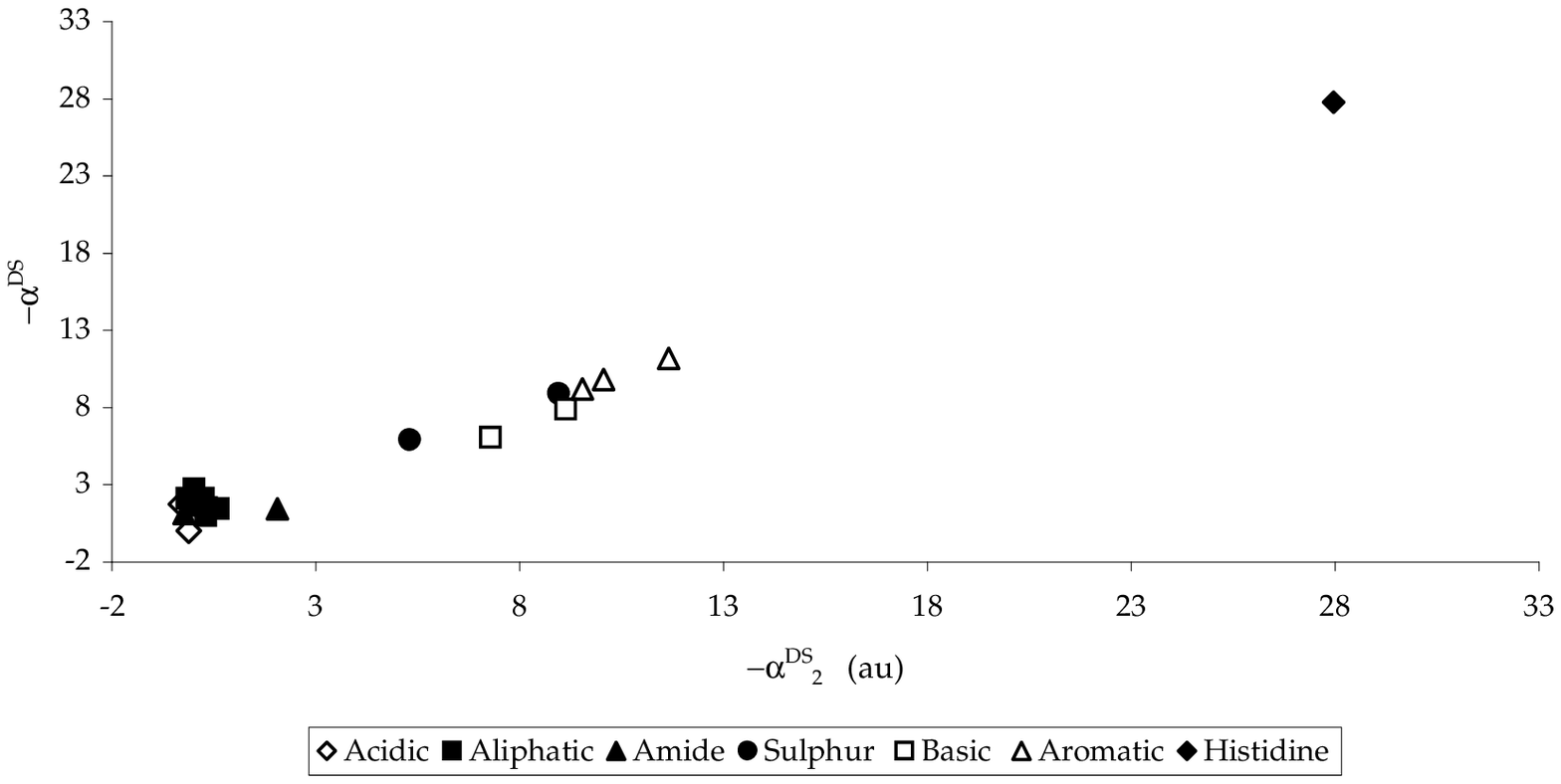}}
\caption{The global dipole-softness polarizability $-\protect\alpha^{DS}$ as
function of the the local dipole-softness polarizabilities of the side chain 
$-\protect\alpha^{DS}_{2}$, calculated from the SCF density obtained with
the 6-311G(d,p) basis set. All values are depicted in au.}
\end{figure}
\newpage 
\begin{figure}[th]
\resizebox{\textwidth}{!}{\includegraphics{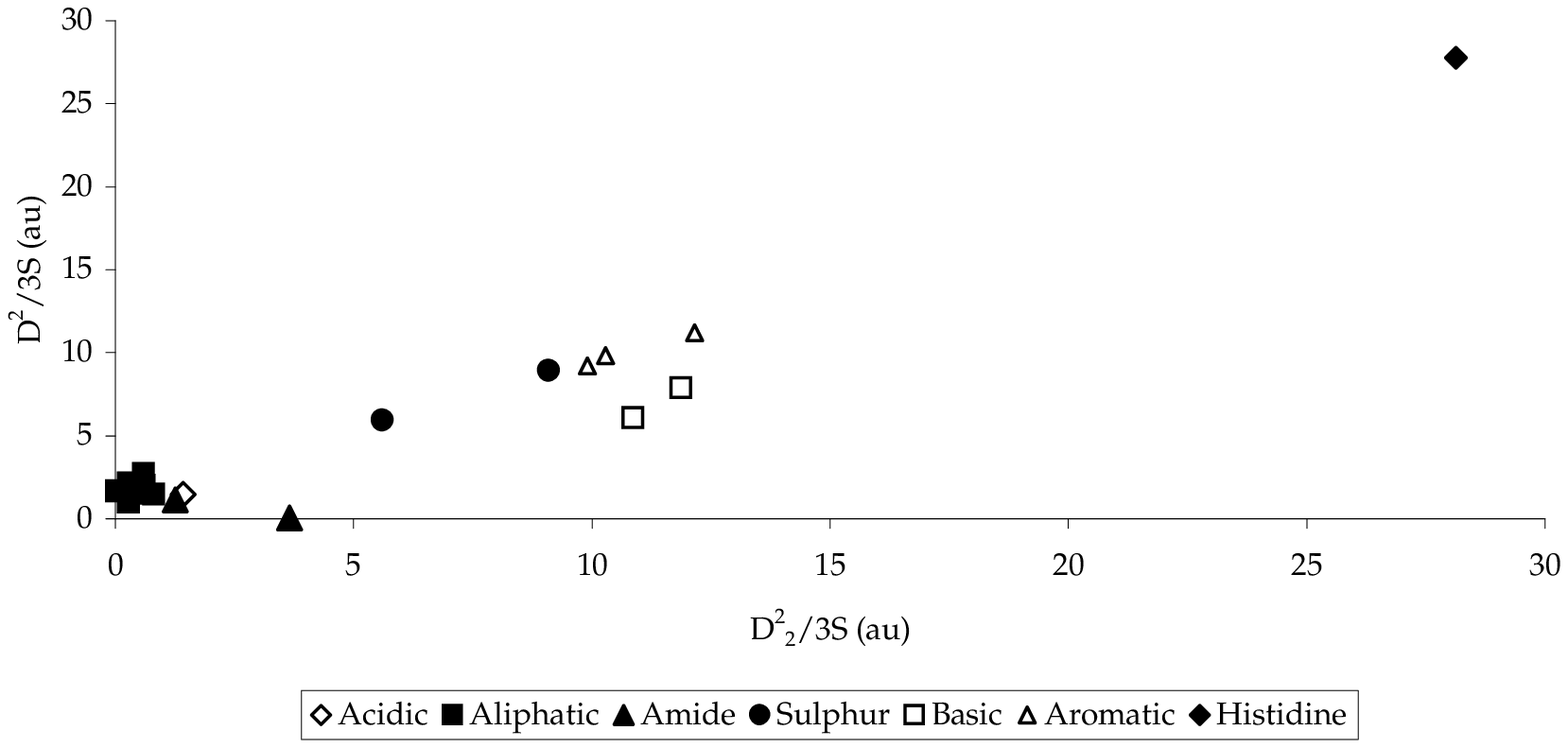}}
\caption{The global dipole-softness polarizability $\frac{D^{2}}{3S}$ as
function of the the local dipole-softness polarizabilities of the side chain 
$\frac{D^{2}_{2}}{3S}$, calculated from the SCF density obtained with the
6-311G(d,p) basis set. All values are depicted in au.}
\end{figure}


\begin{thebibliography}{99}
\bibitem{QuantumSimilarity.ref} M. Karelson, V. S. Lobanov, and A. R.
Katritzky, Chem. Rev. \textbf{96}, 1027 (1996).

\bibitem{softness.ref} W. Yang, R. G. Parr, Proc. Natl. Acad. Sci. USA 
\textbf{82}, 6723 (1985).

\bibitem{Parrbook.ref} R. G. Parr, W. Yang, \textit{Density-Functional
Theory of Atoms and Molecules }(New York: Oxford University Press 1989).

\bibitem{DFT96.ref} P. Senet, J. Chem. Phys. \textbf{105}, 6471 (1996).

\bibitem{Geerlings1.ref} P. Geerlings, F. De Proft, Phys. Chem. Chem. Phys. 
\textbf{10}, 3028 (2008).

\bibitem{Geerlings2.ref} C. Cardenas, F. De Proft, E. Chamorro, P.
Fuentealba, P. Geerlings, J. Chem. Phys \textbf{128}, 034708 (2008).

\bibitem{Kulkarni.ref} B. S. Kulkarni, A. Tanwar, S. Pal, J. Chem. Sciences 
\textbf{119}, 489 (2007).

\bibitem{Hemelsoet.ref} K. Hemelsoet, V. Van Speybroeck, M. Waroquier, Chem.
Phys. Lett \textbf{444}, 17 (2007).

\bibitem{Ayers.ref} P. W. Ayers, Chem. Phys. Lett. \textbf{438}, 148 (2007).

\bibitem{Cohen.ref} M. H. Cohen, A. Wasserman, J. Phys. Chem. A. \textbf{111}%
, 2229 (2007).

\bibitem{Cardenas.ref} P. Fuentealba, E. Chamorro, C. Cardenas, Int. J.
Quant. Chem, \textbf{107}, 37 (2007).

\bibitem{Kar.ref} R. Kar, K. R. S. Chandrakumar, S. Pal, J. Phys. Chem. A. 
\textbf{111}, 375 (2007).

\bibitem{ChemREV.ref} P. Geerlings, F. De Proft, and W. Langenaeker, Chem.
Rev. (Washington, D.C.) \textbf{103}, 1793 (2003).

\bibitem{ChermetteREV.ref} H. Chermette, J. Comput. Chem. \textbf{20}, 129
(1999).

\bibitem{Newbook.ref} Chemical reactivity theory, A Density Functional View,
edited by P.K. Chattaraj, CRC Press (2009).

\bibitem{Pearson86.ref} R. G. Pearson, Proc. Natl. Acad. Sci. U.S.A. \textbf{%
83}, 8440 (1986).

\bibitem{DFT97.ref} P. Senet, J. Chem. Phys. \textbf{107}, 2516 (1997).

\bibitem{CPL.ref} P. Senet, Chem. Phys. Lett. \textbf{275}, 527 (1997).

\bibitem{condens.ref} W. Yang and W. J. Mortier, J. Am. Chem. Soc. \textbf{%
108}, 5708 (1986).

\bibitem{CoulombHole1.ref} P. Senet and M. Yang, J. Chem. Sci. \textbf{117},
411 (2005). \label{CoulombHole1.ref}

\bibitem{CoulombHole.ref} P. Senet, F. Aparicio, J. Chem. Phys. \textbf{126}%
, 145105 (2007). \label{CoulombHole.ref}

\bibitem{Bader.ref} K.E. Laidig and R. F. W. Bader, J. Chem. Phys.\textbf{\
93}, 7213(1990).

\bibitem{Hirshfeld.ref} F. L. Hirshfeld, Theoret. Chim. Acta (Berl.) \textbf{%
44}, 129 (1977).

\bibitem{teixeira-dias.ref} J. J. C. Teixeira-Dias and J. N. Murrell, Mol.
Phys. \textbf{3}, 329 (1970).

\bibitem{Gough.ref} K. M. Gough J. Chem. Phys. \textbf{91}, 2424(1989).

\bibitem{lefevre.ref} R. J. W. Le Fevre, Adv. Phys. Org. Chem. \textbf{3}, 1
(1965).

\bibitem{miller.ref} K. J. Miller, J. Am. Chem. Soc. \textbf{112}, 8533
(1990); \emph{ibid} 8543 (1990).

\bibitem{Rousseau.ref} B. Rousseau, A. Peeters, C. Van Alsenoy, Chem. Phys.
Lett \textbf{324}, 189 (2000).

\bibitem{Bultinck.ref} P. Bultinck, C. Van Alsenoy, P. W. Ayers, R.
Carbo-Dorca, J. Chem. Phys. \textbf{126}, 144111 (2007).

\bibitem{PolarWater.ref} A. Krishtal, P. Senet, M. Yang, C. Van Alsenoy, J.
Chem. Phys. \textbf{125}, 034312 (2006).

\bibitem{PolarMethanol.ref} A. Krishtal, P. Senet, C. Van Alsenoy, J. Chem.
Theory Comput. \textbf{4}, 426 (2008).

\bibitem{Contreras.ref} R. Contreras, F. Fuentealba, M. Galv\'an, P.
P\'erez, Chem. Phys. Lett. \textbf{304}, 405 (1999).

\bibitem{AA1.ref} P. Perez and A. Contreras, Chem. Phys. Lett. 293, 239 1998
.

\bibitem{AA2.ref} J. Melin, F. Aparicio, V. Subramanian, M. Galvan, and P.
Chattaraj, J. Phys. Chem. A \textbf{108}, 2487 2004 .

\bibitem{AA3.ref} Arulmozhiraja, T. Fujii, and G. Sato, Mol. Phys. \textbf{%
100}, 423 2002 .

\bibitem{AA4.ref} A. Baeten, F. De Proft, and P. Geerlings, Int. J. Quantum
Chem. \textbf{60}, 931 1996 .

\bibitem{Resp.ref} J. Callaway, \textit{Quantum Theory of the Solid State.}
San Diego: Academic Press \textbf{1974}.

\bibitem{Mahan.ref} G. D. Mahan and K. R. Subbaswamy, \emph{Local Density
Theory of Polarisability}. Springer, \textbf{1990}. \label{Mahan.ref}

\bibitem{BP.ref} M. Berkowitz, R. G. Parr, J. Chem. Phys. \textbf{88}, 2554
(1988).

\bibitem{Gaussian} M. J. Frisch, G. W. Trucks, H. B. Schlegel, G. E.
Scuseria, M. A. Robb, J. R. Cheeseman, J. A. Montgomery, Jr., T. Vreven, K.
N. Kudin, J. C. Burant, J. M. Millam, S. S. Iyengar, J. Tomasi, V. Barone,
B. Mennucci, M. Cossi, G. Scalmani, N. Rega, G. A. Petersson, H. Nakatsuji,
M. Hada, M. Ehara, K. Toyota, R. Fukuda, J. Hasegawa, M. Ishida, T.
Nakajima, Y. Honda, O. Kitao, H. Nakai, M. Klene, X. Li, J. E. Knox, H. P.
Hratchian, J. B. Cross, V. Bakken, C. Adamo, J. Jaramillo, R. Gomperts, R.
E. Stratmann, O. Yazyev, A. J. Austin, R. Cammi, C. Pomelli, J. W.
Ochterski, P. Y. Ayala, K. Morokuma, G. A. Voth, P. Salvador, J. J.
Dannenberg, V. G. Zakrzewski, S. Dapprich, A. D. Daniels, M. C. Strain, O.
Farkas, D. K. Malick, A. D. Rabuck, K. Raghavachari, J. B. Foresman, J. V.
Ortiz, Q. Cui, A. G. Baboul, S. Clifford, J. Cioslowski, B. B. Stefanov, G.
Liu, A. Liashenko, P. Piskorz, I. Komaromi, R. L. Martin, D. J. Fox, T.
Keith, M. A. Al-Laham, C. Y. Peng, A. Nanayakkara, M. Challacombe, P. M. W.
Gill, B. Johnson, W. Chen, M. W. Wong, C. Gonzalez, and J. A. Pople, \textit{%
Gaussian 03}, Revision C.02 (Gaussian, Inc., Wallingford CT, 2004)

\bibitem{Brabo} C. Van Alsenoy, A. Peeters, J. Mol. Struct (Theochem) 
\textbf{286}, 19 (1993).

\bibitem{Proteins.ref} T. E. Creighton, Proteins, Structures and Molecular
Properties, 2nd ed. Freeman, New York, 1994 .

\bibitem{Politzer2.ref} T. Brinck, J. S. Murray, P. Politzer, J. Chem. Phys. 
\textbf{98}, 4035 (1986) and references therein. \label{Politzer2.ref}

\bibitem{Lambin.ref} See e.g., A. Lucas, L. Henrard and Ph. Lambin, Phys.
Rev. B \textbf{49}, 2888 (1994).
\end{thebibliography}
\end{document}